\documentstyle[psfig,harvard]{elsart}

\newcommand{\ApJL}{Astrophys. J. Lett.}
\newcommand{\ApJ}{Astrophys. J.}
\newcommand{\AandA}{Astron. \& Astrophys.}
\newcommand{\PRL}{Phys. Rev. Lett.}
\newcommand{\PRD}{Phys. Rev. D}
\newcommand{\MNRAS}{Mon. Not. Royal Astron. Soc}
\newcommand{\NA}{New Astron.}

\newcommand{\aut}[2]{{#2.\ #1}}
\newcommand{\refs}[6]{#2, {\bf #3} {#4} (#5)}

\newcommand{\amp}{and }

\newcommand{\da}{d_A}

\newcommand{\cmb}{{\delta I}}

\newcommand{\n}{{\rm n}}
\newcommand{\vecl}{{\bf l}}
\newcommand{\vecla}{{{\bf l}_1}}
\newcommand{\veclb}{{{\bf l}_2}}
\newcommand{\veclc}{{{\bf l}_3}}
\newcommand{\vecld}{{{\bf l}_4}}
\newcommand{\intl}[1]{\int {d^2 {\bf l}_#1 \over (2\pi)^2}}

\newcommand{\bfl}{{\mathbf{l}}}
\newcommand{\dirac}{{\rm D}}

\newcommand{\veck}{{\bf k}}

\newcommand{\lens}{{\rm len}}
\newcommand{\pp}{{\phi\phi}}

\newlength{\tskip}
\newlength{\colwidth}

\newcommand{\beq}{\begin{equation}}
\newcommand{\eeq}{\end{equation}}
\newcommand{\beqa}{\begin{eqnarray}}
\newcommand{\eeqa}{\end{eqnarray}}

\newcommand{\bn}{\hat{\bf n}}
\newcommand{\bm}{\hat{\bf m}}

\newcommand{\rad}{r}    

\newcommand{\len}{\phi}

\begin{document}
\begin{frontmatter}

\title{Lensing Studies with Diffuse Backgrounds}
\author{Asantha Cooray}
\address{
Theoretical Astrophysics, California Institute of Technology,
Pasadena, California 91125\\}


\begin{abstract}
The current weak lensing measurements of the large scale structure are mostly related to statistical
study of background galaxy ellipticities. We consider a possibility to extend lensing studies with
intrinsically unresolved sources and suggest that spatial fluctuations in the integrated diffuse emission 
from these sources can be used for a lensing reconstruction.  Examples of upcoming possibilities include
the diffuse background generated by dusty starburst galaxies at far-infrared wavelengths,
first stars and galaxies in near-infrared wavelengths, and the background related to 21 cm emission by neutral gas in the general
intergalactic medium prior to reionization. While methods developed to extract
lensing information from cosmic microwave background (CMB) temperature and polarization  data can be easily 
modified to study lensing properties using diffuse backgrounds at other wavelengths, we suggest that the lensing
extraction from these backgrounds using higher order non-Gaussian clustering information alone may not be the best approach.
In contrast to CMB anisotropies, reasons for this include the lack of features in the clustering power spectrum
such that the resulting lensing modification to the angular power spectrum of low-redshift diffuse backgrounds, 
at arcminute angular scales, is insignificant. While the use of low redshift backgrounds for lensing studies will be challenging, due to confusing
foregrounds among other reasons, the use of suggested backgrounds will extend the reconstruction of the integrated matter power spectrum out to redshifts
of 15 to 30, and will bridge the gap between current and upcoming galaxy lensing studies out to, at most,
a redshift of a few  and planned weak lensing studies with CMB out to the last scattering surface at a redshift of 1100.
\end{abstract}

\end{frontmatter}

\section{Introduction}

The weak gravitational lensing of background galaxies by the intervening large scale structure is now a well known probe of
physical cosmology (Kaiser 1992; Kaiser 1998; Jain \& Seljak 1997; Hu \& Tegmark 1999; Bartelmann \& Schneider 2001). 
The current weak lensing studies are mostly limited to reduced-shear estimates that involve statistical study of
galaxy ellipticities or shape information reconstructed through expansions related to certain orthogonal basis functions 
(for a recent review, see, Refregier 2003). 
Such a lensing reconstruction, unfortunately, is limited to resolved background sources for which
reliable shape measurements, after accounting for the point spread function, are possible.
While statistics of current and most upcoming weak lensing surveys are expected to be limited by
instrumental constraints, such as issues related to the point spread function, with reasonable improvements on the observational side,
the statistics related to large scale structure lensing studies will eventually be limited by the number of resolved background 
sources. In this limit, to further improve mass reconstruction related to lensing, 
 background objects which are intrinsically point-sources, and, thus, for which shape measurements are impossible, 
 must be included in the analysis. A situation where point-sources dominate resolved source number counts
 is already present in some of the deepest space-based images available to date, including North and South Hubble Deep Fields 
and recent images from the Advanced Camera for Surveys (ACS) on the Hubble Space Telescope.

As one moves to higher source redshifts, beyond what is currently encountered in large scale structure lensing studies,
the importance of  point sources is more evident; The first stars and proto galaxies, at redshifts of order 15 to 30 and higher,
provide an ideal background for lensing studies given the large path length to these background sources.
While the detection of an individual first star is beyond the capabilities of even next generation instruments, the spatial
fluctuations in the integrated diffuse background emission
 from these stars, at least in most optimistic models of their emission, may be within reach (e.g., Cooray et al. 2003).
During the transit to us, this diffuse emission is expected to be affected by the intervening mass distribution via gravitational lensing
and provides a background for a potentially interesting lensing study.

Contrary to background intensities at optical and infra-red wavelengths (Cooray et al. 2003; Knox et al. 2001), 
where the emission can eventually be broken down to point sources, one also encounter backgrounds which are
composed of truly diffuse emission that cannot easily be separated to individual objects.  Examples of such diffuse backgrounds 
are the 21 cm rest wavelength emission from neutral Hydrogen
in the intergalactic medium prior to reionization (Scott \& Rees 1990; Madau et al. 1997; Tozzi et al. 1997; Iliev et al. 2002; Furlanetto et al. 2003) 
and, of course, the cosmic microwave background  (CMB) radiation. The lensing effect on CMB temperature and polarization fluctuations, that
were generated at the last scattering surface (Peebles \& Yu 1970; Sunyaev \& Zel'dovich 1970; Silk 1968; Hu \& Dodelson 2001), 
is now well understood (Seljak 1996; Metcalf \& Silk 1997; Hu 2000)
and various techniques have been developed to extract lensing information from high resolution CMB temperature and polarization maps 
(Zaldarriaga \& Seljak 1999; Benabed et al. 2001; Guzik et al. 2000; Hu 2001; Hu \& Okamoto 2002; Cooray \& Kesden 2003; Hirata \& Seljak 2003).
These techniques make use of the fact that at the last scattering surface, where primary CMB temperature and polarization fluctuations are generated, 
these anisotropies follow a Gaussian distribution, while with gravitational lensing deflections when propagating to us,
the fluctuations pattern on the sky is non-Gaussian (Bernardeau 1997; Hu 2001; Zaldarriaga 2000; kesden et al. 2002). 
The lensing extraction techniques make use of
 measurements related to certain non-Gaussian aspects, such as a collapsed form of the trispectrum (Zaldarriaga 2000; Hu 2001),
which is related to a filtered version of the squared-temperature power spectrum (Cooray \& Kesden 2003), to directly estimate and reconstruct the
deflection angle or the projected lensing potential power spectrum.

In the case of CMB, the study of lensing and its extraction is usually considered under a perturbative approach with the deflection angle as a
small change to the anisotropy on the sky towards a given direction (Hu 2000; Goldberg \& Spergel 1998; Cooray \& Hu 1998; Zaldarriaga 2000). 
Under such an approach, for most practical purposes, the lensing effect can be
calculated up to the second order in the deflection angle or first order in the lensing potential power spectrum. 
Since the rms deflection angle for a photon propagating from a redshift of 1100 is of order a few arcminute, a resulting 
complication here is that perturbative description of the lensing effect breaks down at small angular scales at and below the rms deflection angle.
This, however, is not a major concern for the CMB lensing description since there is no intrinsic small angular scale power at these
same angular scales due to the presence of significant damping.
On the other hand, when compared to CMB temperature fluctuations,  diffuse backgrounds 
at lower redshifts are expected to show substantial clustering power at arcminute angular scales 
and the perturbative approach just to the first order in the deflection power spectrum may not be an adequate description.
Another important difference, when compared to CMB anisotropies that were generated at the last scattering surface with narrow range in redshift, 
is that background sources may be distributed over a wide range in redshift and this distribution must be properly accounted as well.

While there are complications from foregrounds and other confusions, we suggest that 
 techniques developed to extract lensing information from CMB data can also be extended and applied at other
wavelengths for lensing studies.   In the case of CMB anisotropies, 
there are substantial modifications to the anisotropy angular power spectrum from the lensing effect due to the
acoustic peak structure and the presence of a damping tail such that lensing transfers power from large scales to arcminute scales
where the lensing effect is easily detectable. When compared to a lensing study with CMB, 
the lensing information one can extract from low redshift diffuse backgrounds is significantly limited. As we will discuss, this 
results from the fact that,
at arcminute angular scales, the lensing modification to the power spectrum is minor and the non-Gaussianities generated are
relatively smaller. This, again, is due to the lack of features in the clustering power spectrum such that the modification to
the clustering pattern, at arcminute angular scales of interest for future observations, remains the
same under gravitational lensing. 
The lack of significant differences between the intrinsic and lensed power spectra, thus, limits the use of statistics such as
clustering variance for lensing purposes.

To extract lensing information better, one require arcsecond scale clustering information, though at such small angular scales, 
a substantial number of sources that contribute to the diffuse background will begin to be resolved.
In the limit that most or all sources are resolved to point-like objects, instead of shear reconstruction based on source shapes, 
statistics related to the distribution of the point sources can also be used for a shear estimation.
As an example, instead of extracting information from the  non-Gaussian clustering information, one can make use of the anisotropy
of clustering generated by foreground lensing. This anisotropy can be used as an estimator of shear and can be approached in the same manner
ellipticity information from galaxy shapes are used for an estimate of shear. Such a study has
already been attempted towards known regions of mass concentrations,
such as galaxy clusters (van Waerbeke et al. 1997). Similar studies
can be extended to large angular scales and the large scale structure as a whole when carefully selected backgrounds
are selected for lensing reconstruction so as to minimize confusions resulting from foregrounds.
Another aspect is the change resulting from
lensing magnification and  several suggestions have already been made to include unresolved sources in weak lensing studies through 
 variations associated with number counts and, when measurable, source sizes (Jain 2002; Pen 2003). 

The paper is organized as follows. In the next section, we discuss the lensing effect on diffuse backgrounds and extend calculations
related to weak lensing effect on CMB anisotropies. As an example, 
we discuss the lensing modification of the
near-IR anisotropy fluctuations due to first stars following calculations, related to spatial clustering discussed in Cooray et al. (2003).
We discuss the extent to which lensing information can be extracted  from this and other backgrounds and a comparison to the lensing
extraction with CMB data. We also consider potential biases and systematic effects for a lensing reconstruction and conclude with a summary.

\section{Calculation}

\subsection{Lensing Effect on Clustering of Diffuse Backgrounds}

In order to derive the weak lensing effect on angular clustering properties of any diffuse background, we follow a calculation similar to that
introduced for the lensing extraction from CMB temperature anisotropies (e.g., Hu 2001; Seljak \& Zaldarriaga 1999; Cooray \& Kesden 2003)
 and make use of the flat sky approximation. When compared to CMB, there is one important
difference in that one should properly account for the broad redshift distribution of sources that contribute to the background light, instead of the 
assumption related to a narrow source redshift distribution. Thus, we break the background intensity as 
\begin{equation}
I_{\rm tot}(\bn) =\int_0^{\infty} d\rad \cmb(\rad \bn, \rad) \, ,
\label{eqn:total}
\end{equation}
where $\cmb(\rad \bn,\rad)$ is the fractional contribution to the total emission
as a function of the radial distance or look-back time, from the observer:
\begin{equation}
\rad(z) = \int_0^z {dz' \over H(z')} \,,
\end{equation}
when the expansion rate for adiabatic CDM
cosmological models with a cosmological constant is given by
\begin{equation}
H^2 = H_0^2 \left[ \Omega_m(1+z)^3 + \Omega_K (1+z)^2
+\Omega_\Lambda \right]\,.
\end{equation}
Here, $H_0$ can be written as the inverse
Hubble distance today $cH_0^{-1} = 2997.9h^{-1}$ Mpc.
We follow the conventions that in units of the critical
density $3H_0^2/8\pi G$, the contribution of each component is denoted
$\Omega_i$, $i=c$ for the CDM, $b$ for the baryons,
$\Lambda$ for the cosmological constant.
We also define the auxiliary quantities
$\Omega_m=\Omega_c+\Omega_b$ and $\Omega_K=1-\sum_i \Omega_i$,
which represent the matter density and the contribution of
spatial curvature to the expansion rate respectively.
Note that, though we discuss a general derivation of lensing effect on background source clustering,
we show results for the currently  favorable $\Lambda$CDM cosmology with
$\Omega_b=0.05$, $\Omega_m=0.35$, $\Omega_\Lambda=0.65$ and $h=0.65$.

Weak lensing deflects the path of background photons resulting in a remapping of the observed anisotropy pattern on the sky, such that the
fractional contribution from each redshift is modified as
\begin{eqnarray}
&&\tilde \cmb(\rad \bn,\rad)  =   \cmb[ \rad \{\bn + \nabla\len(\rad \bn,\rad)\},\rad] \nonumber\\
        & \approx &
\cmb(\rad \bn,\rad) + \nabla_i \len(\rad \bn,\rad) \nabla^i \cmb(\rad \bn,\rad) + {1 \over 2} \nabla_i \len(\rad \bn,\rad) \nabla_j \len(\rad \bn,\rad) \nabla^{i}\nabla^{j} \cmb(\rad \bn, \rad)
+ \ldots  \; .
\label{eqn:pert}
\end{eqnarray}
Here, $\cmb(\rad \bn,\rad)$ is the unlensed fractional component of the background light from a distance of $\rad$, 
$\tilde \cmb(\rad \bn, \rad)$ is the same contribution  when affected by the gravitational lensing deflections during the transit, and
$\nabla\len(\rad \bn, \rad)$ represents the lensing deflection angle for a photon propagating from a distance of $\rad$. 
The second line of Eq.~\ref{eqn:pert} treats the lensing deflection as a perturbative parameter. Typical lensing calculations
related to CMB (e.g., Hu 2000), treat lensing effect on the angular power
spectrum to the second order in this perturbation expansion, while lensing extraction techniques
related to CMB consider information from the first order correction (e.g., Seljak \& Zaldarriaga 2000; Hu 2001). 
 Such an approach works well since CMB has no
intrinsic anisotropy at small angular scales below few arcminutes, though the typical deflection angle of a photon
from the last scattering surface is $\sim$ 2.5 arcminutes.
The accuracy to which lensing can be extracted from CMB under this perturbative approach is discussed in Hirata \& Seljak (2003).
Here, for low redshift diffuse backgrounds, the perturbation expansion up to second order in the deflection angle is
not adequate to fully  account for the lensing effect on the clustering power spectrum. We extend the calculation explicitly to higher order
and present an exact derivation of the lensed power spectrum which should be valid even at small angular scales
where the lensing deflection angle is of the same order. To properly evaluate the lensed contribution, however, one must evaluate an infinite series
of integrals which is numerically exhaustive; here, in presenting numerical results, we only
 consider the lensed power spectrum up to the fourth order in deflection angle or the second order in the deflection-angle power spectrum.

In equation~\ref{eqn:pert}, where $\nabla \phi$ is the deflection angle, 
$\phi$ is a radial projection of the gravitational potential, $\Phi$ (see, e.g. Kaiser 1992), for a source at a distance $\rad_s$
\begin{eqnarray}
\phi(\rad_s \bm, \rad_s)&=&- 2 \int_0^{\rad_s} d\rad' 
\frac{\da(\rad_s-\rad')}{\da(\rad_s)\da(\rad')} \Phi (\rad' \hat{{\bf m}},\rad') \, ,
\label{eqn:lenspotential}
\end{eqnarray}
where the comoving angular diameter distance is
\begin{equation}
\da = H_0^{-1} \Omega_K^{-1/2} \sinh (H_0 \Omega_K^{1/2} \rad)\, .
\end{equation}
Note that as $\Omega_K \rightarrow 0$, $\da \rightarrow \rad$.

Taking  the Fourier transform, as appropriate for a flat-sky, we write the modification to the fraction contribution arising at
a distance $\rad$ as
\begin{eqnarray}
\tilde \cmb(\vecla)
&=& \int d \bn\, \tilde \cmb(\bn) e^{-i \vecla \cdot \bn} \nonumber\\
&=& \cmb(\vecla) - \intl{1'} \cmb(\vecla') L(\vecla,\vecla')\,,
\label{eqn:thetal}
\end{eqnarray}
where
\begin{eqnarray}
\label{eqn:lfactor}
&&L(\vecla,\vecla') \equiv \len(\vecla-\vecla') \, (\vecla - \vecla') \cdot \vecla'
-{1 \over 2} \intl{1''} \len(\vecla'') \nonumber \\ 
&\times& \len(\vecla - \vecla' - \vecla'') \, (\vecla'' \cdot \vecla')
(\vecla - \vecla' - \vecla'')\cdot \vecla' + {1 \over 6} \intl{1''} \intl{1'''}\len(\vecla'') \len(\vecla''') \nonumber \\ 
&\times& \len(\vecla - \vecla' - \vecla'' -\vecla''') \, (\vecla'' \cdot \vecla') (\vecla''' \cdot \vecla') \,
(\vecla - \vecla' - \vecla'' -\vecla''')\cdot \vecla' \,. + .... \nonumber \\
\end{eqnarray}
Here, for simplicity, we have dropped the explicit dependence on the radial distance, $\rad$, in $\cmb$ and $\phi$,
and have expanded the correction to the third order in $\phi$.

We define the power spectrum of the total intensity fluctuation field, in the flat
sky approximation, following the usual way
\begin{eqnarray}
\left< I(\bfl) I(\bfl')\right> &\equiv&
        (2\pi)^2 \delta_\dirac(\vecl+\vecl')  C_l\,,
\end{eqnarray}
where $\delta_D$ is the Dirac delta function. Similarly, the power spectrum of the lensed fluctuation field, with $I(\bfl)$ replaced by
$\tilde I(\bfl)$, is $\tilde C_l$.

Note that in the absence of lensing, the observed anisotropy 
power spectrum is simply the sum of anisotropy contributions over the normalized source redshift distribution such that $C_l = \int dr 
C_l^{\delta I}$.  The observed angular power spectrum, however,
 consists of both the unlensed intensity and a perturbative correction related to the lensing
effect. To calculate the final lensed-clustering power spectrum, we first substitute 
Eq.~\ref{eqn:thetal} in Eq.~\ref{eqn:total} and reintroduce the dependence on distance. 
The lensing contribution to the power spectrum is calculated following approaches used to describe lensing effect on CMB anisotropies (e.g., Hu 2000)
and after some straight forward calculations, we obtain, to the fourth order in $\phi$
 or second order in $C_l^{\phi}$:
\begin{eqnarray}
&&\tilde C_l  =                C_l  - l^2 \int d\rad  R(r) C_l^{\delta I} \left[1 - \frac{l^2}{2} R(r)\right]              \nonumber \\
  &+& \intl{1} \int d\rad C^{\phi}_{l_1}(\rad) C_{| \vecl - \vecl_1|}^{\delta I} 
    [(\vecl - \vecl_1)\cdot \vecl_1]^2 \left[1 - |\vecl-\vecl_1|^2 R(r) \right] \nonumber \\
    &+& {1 \over 2}\intl{1} \intl{2} \int d\rad C^{\phi}_{l_1}(\rad) C^{\phi}_{l_2}(\rad) C_{|\vecl-\vecl_1-\vecl_2|}^{\delta I} 
[(\vecl - \vecl_1 -\vecl_2)\cdot \vecl_1]^2 [(\vecl - \vecl_1 -\vecl_2)\cdot \vecl_2]^2 \, , \nonumber \\
\label{eqn:ttflat}
\end{eqnarray}
where $C_l^{\delta I}$ is the fractional contribution to the intensity 
fluctuation power spectrum as a function of the radial distance (see, Section 2.2) and
\begin{equation}
R(r)={1 \over  4\pi} \int dl l^3 C_l^{\phi}(\rad) \, .
\end{equation}
The cumulative variance of intensity fluctuations is conserved
under lensing such that $\sigma^2 \equiv \int d^2\vecl/(2\pi)^2 \tilde C_l = \int d^2\vecl/(2\pi)^2  C_l$; this is equivalent to the
fact that lensing does not create or destroy power but rather results in a redistribution such that surface brightness remains
conserved. Under lensing, however, the filtered variance of fluctuations as a function of angular scale is different
such that $\sigma^2(\theta) \equiv \int d^2\vecl/(2\pi)^2 \tilde C_l W^2(l \theta)$, where $W(l \theta)$ is the Fourier transform of the filter --- such as top-hat, Gaussian etc. --- is different from that expected
from the unlensed clustering power spectrum, $\int d^2\vecl/(2\pi)^2  C_l W^2(l \theta)$. While the modification to variance resulting
from lensing can be used for a reconstruction of convergence (Pen 2003), such a measurement only works out accurately if the differences between 
$\tilde C_l$ and $C_l$ is significant, as a function of angular scale or multipole $l$, under gravitational lensing.
While this is the case for arcminute scale power spectrum of CMB anisotropies, we will show that the difference between 
intrinsic- and lensed-power spectra of diffuse backgrounds is insignificant and will complicate a straight forward lensing reconstruction based on
the variance alone.

In the case of CMB anisotropies, in Eq.~1, $\cmb(\rad \bn,\rad)=I^{\rm cmb}(\bn) \delta_D(\rad-\rad_0)$ where $\rad_0$ is the distance to the
surface of last scattering and $I^{\rm cmb}(\bn)$ is the total CMB intensity
 such that for spatial fluctuations, $C_l^{\delta I} = C_l^{\rm cmb}\delta_D(\rad-\rad_0)$ with the primordial anisotropy power
spectrum given by $C_l^{\rm cmb}$. This  assumption simplifies  Eq.~\ref{eqn:ttflat} to  a result
valid for any background source of narrow width in distance or in redshift space, as
\begin{eqnarray}
&&\tilde C_l^{\rm CMB}  =                C_l^{\rm CMB} \left[1 - l^2 R(\rad_0) +\frac{l^4}{2} R^2(\rad_0)\right] \nonumber \\
  &+& \intl{1} C^{\phi}_{l_1}(\rad_0) C_{| \vecl - \vecl_1|}^{\rm CMB} 
  [(\vecl - \vecl_1)\cdot \vecl_1]^2 \left[1 - |\vecl-\vecl_1|^2 R(\rad_0) \right] \nonumber \\
  &+& {1 \over 2}\intl{1} \intl{2} C^{\phi}_{l_1}(\rad_0) C^{\phi}_{l_2}(\rad_0) C_{|\vecl-\vecl_1-\vecl_2|}^{\rm CMB} 
[(\vecl - \vecl_1 -\vecl_2)\cdot \vecl_1]^2 [(\vecl - \vecl_1 -\vecl_2)\cdot \vecl_2]^2 \, . \nonumber \\
\label{eqn:cmbflat}
\end{eqnarray}
This equation is equivalent to that in Hu (2000) for weak lensing effect on CMB anisotropy power spectrum
when one drops higher order terms which are proportional to $[C_l^{\phi}]^2$, $R^2$ and
$C_l^{\phi} R$. As we will discuss, to describe the lensing effect related to CMB, these higher order terms are not crucial.

When one extends the perturbative calculation further, the power spectrum can be represented as a sum of a series of integrals
such that
\begin{eqnarray}
&&\tilde C_l^{\rm CMB}  =                C_l^{\rm CMB} \sum_{i=0}^{\infty} (-1)^i \frac{l^{2i}R(r_0)^i}{i!} \nonumber \\
  &+& \intl{1} C^{\phi}_{l_1}(\rad_0) C_{| \vecl - \vecl_1|}^{\rm CMB} 
    [(\vecl - \vecl_1)\cdot \vecl_1]^2 \sum_{i=0}^{\infty} (-1)^i \frac{|\vecl - \vecl_1|^{2i}R(\rad_0)^i}{i!} \nonumber \\
&+& {1 \over 2} \intl{1} \intl{2}  C^{\phi}_{l_1}(\rad_0) C^{\phi}_{l_2}(\rad_0) C_{|\vecl-\vecl_1-\vecl_2|}^{\rm CMB} 
[(\vecl - \vecl_1 -\vecl_2)\cdot \vecl_1]^2 [(\vecl - \vecl_1 -\vecl_2)\cdot \vecl_2]^2  \nonumber \\
&& \quad \quad \times \sum_{i=0}^{\infty} (-1)^i \frac{|\vecl - \vecl_1 - \vecl_2|^{2i}R(r_0)^i}{i!} + ... +   \nonumber \\
&+& {1 \over n!} \intl{1} ... \intl{n}  C^{\phi}_{l_1}(\rad_0) ... C^{\phi}_{l_n}(\rad_0) C_{|\vecl-...-\vecl_n|}^{\rm CMB}
[(\vecl - ... -\vecl_n)\cdot \vecl_1]^2 ... [(\vecl - ... -\vecl_n)\cdot \vecl_n]^2 \nonumber \\
&& \quad \quad \times \sum_{i=0}^{\infty} (-1)^i \frac{|\vecl - ... - \vecl_n|^{2i}R(r_0)^i}{i!} \, , 
\label{eqn:seriesflat}
\end{eqnarray}
where the last term is a generalization of this series to the n-th term. The summations over $R(r_0)$ in each of these terms
 can be simplified analytically and we obtain
\begin{eqnarray}
&&\tilde C_l^{\rm CMB}  =         C_l^{\rm CMB} e^{-l^2 R(r_0)}\nonumber \\
  &+& \intl{1} C^{\phi}_{l_1}(\rad_0) C_{| \vecl - \vecl_1|}^{\rm CMB}
    [(\vecl - \vecl_1)\cdot \vecl_1]^2 e^{-|\vecl - \vecl_1|^{2}R(r_0)}\nonumber \\
&+& {1 \over 2} \intl{1} \intl{2}  C^{\phi}_{l_1}(\rad_0) C^{\phi}_{l_2}(\rad_0) C_{|\vecl-\vecl_1-\vecl_2|}^{\rm CMB}
[(\vecl - \vecl_1 -\vecl_2)\cdot \vecl_1]^2 [(\vecl - \vecl_1 -\vecl_2)\cdot \vecl_2]^2  \nonumber \\
&& \quad \quad \times e^{-|\vecl - \vecl_1 - \vecl_2|^{2}R(r_0)} + ... +  \nonumber \\
&+& {1 \over n!} \intl{1} ... \intl{n} C^{\phi}_{l_1}(\rad_0) ... C^{\phi}_{l_n}(\rad_0) C_{|\vecl-...-\vecl_n|}^{\rm CMB}
[(\vecl - ... -\vecl_n)\cdot \vecl_1]^2 ... [(\vecl - ... -\vecl_n)\cdot \vecl_n]^2 \nonumber \\
&& \quad \quad \times e^{-|\vecl - ... - \vecl_n|^{2}R(r_0)} \, .
\label{eqn:totalflat}
\end{eqnarray}
Though this expression is useful since it presents the exact calculation related to how gravitational lensing
modifies the clustering power spectrum, when numerically evaluating the lensed power spectrum with a 
limited number of terms in this series, one cannot simply evaluate 
the first few terms and ignore the rest. To evaluate the contribution accurately, one should include terms which are same order in $C_l^{\phi}$ and $R$;
For example, to the second order in $C_l^{\phi}$, contributions include
three terms from  the first line in Eq.~14, two terms from the second and a term from the third, when exponentials in each of the above lines
 are expanded. 

Note that the lensing effect only modifies the clustering pattern of an intrinsically clustered background. If the spatial distribution of sources
responsible for the diffuse background is simply Poisson, then one expects a shot-noise like angular power spectrum for
fluctuations in the intensity such that $C_l^\cmb$ is a constant. In this case, Eq.~\ref{eqn:ttflat} can be simplified  
and leads to $\tilde C_l = C_l$  such that spatial fluctuations, after lensing, remains a Poisson distribution with the same shot-noise.
This is again consistent with the fact that lensing does not create or destroy power but only results in a modification to how power is
 distributed.

The same can also be deduced from Eq.~\ref{eqn:totalflat} by taking into account the exact calculation instead of the approximate one in
Eq.~\ref{eqn:ttflat}. In the case of an intrinsic Poisson-clustered power spectrum, or a
smooth or slowly varying intrinsic power spectrum, one can set $C^{\rm CMB}_{|\vecl -...-\vecl_n|} = C^{\rm CMB}_l$ 
with the substitution
that $|\vecl -...-\vecl_n| \sim l$. 
In this case, one can write Eq.~\ref{eqn:totalflat} as
\begin{eqnarray}
\tilde C_l^{\rm CMB}  &=&         C_l^{\rm CMB} e^{-l^2 R(r_0)} \Big\{1 + ... \nonumber \\ 
&+& {1 \over n!} 
\intl{1} ... \intl{n} C^{\phi}_{l_1}(\rad_0) ... C^{\phi}_{l_n}(\rad_0) [\vecl \cdot \vecl_1]^2 ... [\vecl \cdot \vecl_n]^2 \Big\} \, , \nonumber \\
\end{eqnarray}
which simplifies as $\tilde C_l^{\rm CMB}  =         C_l^{\rm CMB}$ since the series within curly brackets is simply $e^{l^2 R(r_0)}$.
This is important since as we discuss later, lensing only modifies the clustering pattern of background sources 
if the distribution has distinct signatures, such that the angular power spectrum show features. In the case of a smooth 
distribution, which may also be Poisson, lensing does not modify the clustering pattern significantly.
In this sense, most shot-noise dominated backgrounds such as radio sources
and galaxy clusters through the SZ effect at arcminute scale CMB experiments, will not be modified by gravitational lensing. 
We find a similar situation for some what higher redshift backgrounds at IR and other wavelengths.
The only lensing modification is then associated with magnification changes to  source number counts (Tegmark \& Villumsen 1997)
such that, down to a certain flux limit, more sources are magnified towards regions of high mass concentrations
and the extent to which this is important depends strongly on the distribution of number counts as a function of flux.
For truly diffuse backgrounds such as CMB and 21cm emission from the general IGM prior to reionization, such an effect is not expected.

\begin{figure}[t]
\centerline{
\psfig{file=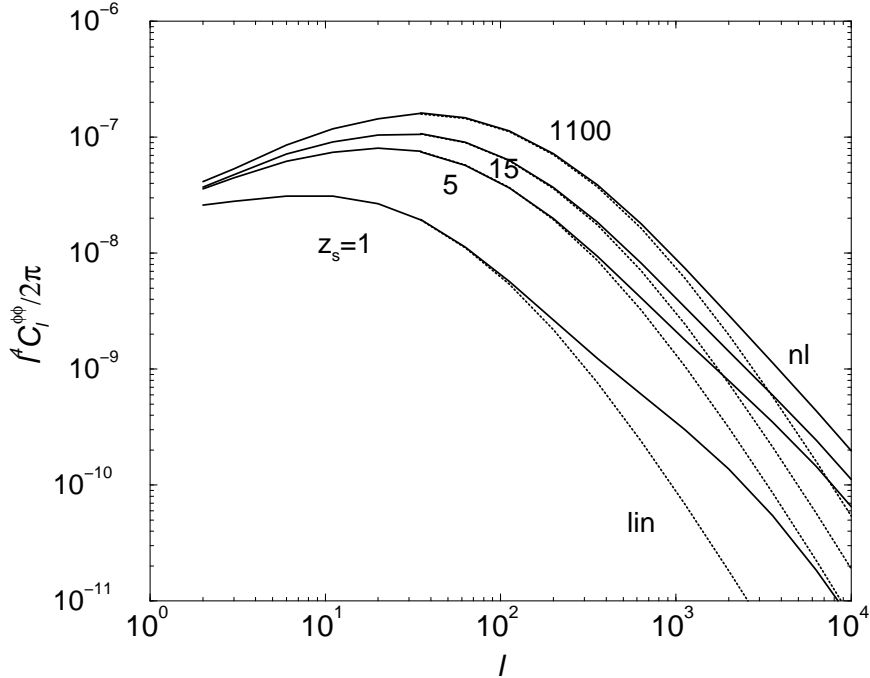,width=4.5in,angle=-90}}
\caption{Angular power spectra of the lensing deflections as a function of the source redshift $z_s$. The solid lines show the
 fully non-linear contribution to the power spectra, while the dotted lines show the contribution under linear theory power
spectrum. While low redshift lensing requires a fully non-linear treatment, for sources at redshifts of 15 and higher, non-linearities
do not affect lensing deflections at angular scales at least of few arcminutes and below ($l \sim 10^3$).}
\label{fig:clphi}
\end{figure}

In Eq.~\ref{eqn:ttflat} and after,
$C^{\phi}_{l}(\rad_s)$ is the power spectrum of lensing potentials, for a source at a distance of $\rad_s$.
In terms of the three-dimensional matter power spectrum, $P(k)$, we can write this power spectrum of lensing potentials as
\begin{eqnarray}
C_l^{\phi}(\rad_s) &=& \frac{2}{\pi} \int k^2\, dk P(k)
\int_0^{\rad_s} d\rad W^\lens(k,\rad_1) j_l(k\rad_1) \int_0^{\rad_s} d\rad_2 W^\lens(k,\rad_2) j_l(k\rad_2) \, ,
\label{eqn:cllens}
\end{eqnarray}
where
\begin{eqnarray}
W^\lens(k,r) &=& -3\Omega_m \left(\frac{H_0}{k}\right)^2 \frac{G(\rad)}{a}
\frac{\da(\rad_s-\rad)}{\da(\rad)\da(\rad_s)} \, .
\end{eqnarray}
In terms of the well known convergence, $\kappa$, power spectrum, that define the integrated surface density of mass out to the background 
distance $\rad_s$, one can write $C_l^{\kappa}= l^4/4C_l^\phi$. 
In calculating the matter power spectrum, we use the fitting formulae of Eisenstein \& Hu (1999)
to evaluate the transfer function for CDM models and set the normalization to match both COBE and
clusters (Bunn \& White 1999; Viana \& Liddle 1999).  Note that in linear theory, the power
spectrum can be scaled in time, $P(k,r)=G^2(r)P(k,0)$, using the the growth function  (Peebles 1980).
In the non-linear regime, one can use prescriptions such as the fitting function by Peacock \& Doods (1996) or halo-based approaches (Cooray \& Sheth 2002)
to calculate the fully non-linear density field power spectrum.

We illustrate these power spectra as a function of $z_s$, the source redshift, in Fig.~1.  Note that for sources at low redshifts, $z_s \sim 1$,
non-linear corrections to the matter power spectrum is important since lensing signal from degree angular scales and below are dominated
by non-linearities. As one moves to a higher source redshift, non-linear corrections move to smaller angular scales
such that for sources at redshifts 15 and higher, non-linear corrections are no longer important as lensing statistics
at arcminute scales are fully described by linear fluctuations in the intervening matter density field. For CMB photons,
the rms deflection angle, $\theta_{\rm rms} = [\int d^2\vecl/(2\pi)^2 l^2 C_l^\phi]^{0.5}$, is 2.6$'$ while for a source at
a redshift of 15, $\theta_{\rm rms} \sim 2'$. The fact that the typical deflection angle is of order few arcminutes, require that clustering
modifications not be studied just to the first order perturbative correction and that higher order terms be included.

In Fig.~2, we show the CMB anisotropy power spectrum and the resulting modification related to the gravitational lensing effect. Note that the
lensing effect moves power from large angular scales, where acoustic peaks are found, to small angular scales (Seljak 1996; Hu 2000). This results in
an increase of power at arcminute angular scales where the damping tail of the anisotropy power spectrum is found. We also show the correction
related to second order in the deflection angle power spectrum. The resulting modification  related to this additional higher order correction is at the
level of 20\%, at most, though it becomes significant as $l \rightarrow 10^4$, suggesting that, at small angular scales, the perturbative approach
may be suspect to describe the lensing effect on CMB. Still, the approach considered in certain calculations (e.g., Hu 2000) is
adequate since CMB anisotropies have no significant power at small angular scales and higher order terms, which are important at small angular scales,
have a decreasing significance. For precision cosmology with arcminute angular-scale CMB power spectrum measurements,
 however, it may be useful to include higher order terms since  they are above the cosmic variance limit of the
power spectrum measurements. In addition to these higher order corrections, there are other assumptions that can complicate the interpretation of
CMB lensing data. For example,  the CMB power spectrum contains an anisotropy contribution from low redshifts
 through secondary effects such as the integrated Sachs-Wolfe effect (Sachs \& Wolfe 1967) and the Sunyaev-Zel'dovich effect (Sunyaev \& Zel'dovich 1980).
Since these contributions are also lensed, simply using lensed power spectrum with a narrow source at $\rad_0$ for cosmological 
parameter interpretation may be problematic. This issue is beyond the purpose of the present paper and we will leave it for a future study.

As we will now discuss, due to the smoothness of spatial fluctuations generated by low redshift diffuse backgrounds at IR wavelengths and
at 21 cm rest-wavelength by neutral Hydrogen rich IGM  prior to reionization, the lensing
modifications, at arcminute angular scales, are not significant. This limits their ability for a lensing reconstruction in the same manner as
approached for CMB.

\subsection{Diffuse Background Clustering Power Spectrum}

Since gravitational lensing deflections modify the clustering power spectrum of anisotropies in the diffuse background, we first consider an
estimate of the intrinsic spatial clustering signal. In the introduction, we identified several potential backgrounds that may become important for
weak lensing studies including that due to first stars and galaxies in near-IR wavelengths and the 21 cm background by the general IGM
prior to reionization. Instead of making separate estimates of their clustering signals here, we consider a general calculation under
the assumption that background sources, either in the form of first stars or neutral Hydrogen, 
trace linear fluctuations with a bias factor that
can be calculated based on the halo mass in which these sources are expected. In the case of first stars, such halos are likely to have
temperatures of 200 K, if molecular Hydrogen cooling is allowed, or 10$^4$ K, or more. 
In the case of diffuse emission related to, say 21 cm, from the neutral
hydrogen prior to reionization, we expect neutral Hydrogen to be present in all halos with no  cut off at the low end of the mass distribution.
In this case, it is likely that the neutral Hydrogen distribution is tracing the linear density field directly such that the bias
factor is close to unity.

The clustering of any sources at redshifts between 10 and 30 can be calculated following standard approaches (Cooray et al. 2003).
The contribution to the background intensity, say at a certain wavelength and towards a 
direction $\bn$,  can be written as a product of the mean emissivity and its fluctuation
\begin{equation}
I(\bn) = \int_0^{\infty} dr  a(r) \bar{j}(r) \left[1+\frac{\delta j(r\bn,r)}{\bar{j}(r)}\right] \, ,
\end{equation}
where $r$ is the comoving radial distance (or conformal time) and $\bar{j}(r)$ is the mean emissivity per comoving
unit volume as a function of distance. Note that the integrand can now be identified with $\delta I(r\bn,r)$ in Eq.~1.
In order to calculate spatial fluctuations related to the emissivity, we assume $\delta j(r\bn,r)/\bar{j}(r)$ trace
fluctuations in the source density field, $\delta_s=\delta \rho_s/\bar{\rho}_s$ such that, in Fourier space,
\begin{equation}
\frac{\delta j(\veck,z)}{\bar{j}(z)}= \delta_s(\veck,z) \, .
\end{equation}
The density field fluctuations of emitters are defined by the three dimensional power spectrum, which we define as
\begin{equation} 
\langle \delta_s(\veck,z) \delta_s(\veck',z) \rangle = (2\pi)^3 \delta_D(\veck+\veck')P_{ss}(k,z) \, .
\end{equation}
To calculate the source power spectrum, as discussed, we scale the linear clustering power spectrum by a bias factor.
While in the case of 21 cm emission from neutral gas the source bias is expected to mostly unity, for sources that
form in halos above a certain mass cut-off, $M_{\rm cut}$, the bias factor is simply given by 
\begin{equation}
\langle b_M \rangle  = \frac{\int_{M_{\rm cut}}^\infty dm \; m\; b(m) n(m)}{\int_{M_{\rm cut}}^\infty dm\; m\;  n(m)} \, ,
\end{equation}
where $b(m)$ is the halo bias (Mo \& White 1996; Mo et al. 1997) with respect to the density field and $n(m)$ is the mass function (Press \& Schechter 1974).

Using the Limber approximation (Limber 1954), the angular power spectrum for a distribution of sources that trace a
three-dimensional power spectrum $P_{ss}(k)$, when projected on the sky, is given by
\begin{equation}
C_l  = \int dr \frac{a^2(r)}{d_A^2}
\bar{j}^2(z)  G^2(r) \langle b_M(r) \rangle^2 P^{\rm lin}\left(k=\frac{l}{d_A},r=0\right) \, ,
\end{equation}
and the integrand is $C_l^{\delta I}(\rad)$ in Eq.~\ref{eqn:ttflat}.

The power spectrum of diffuse emission related to sources in IR is illustrated in Fig.~3. The calculation related to $\bar{j}(z)$ in this case
follows that of Cooray et al. (2003) and we make use of the most optimistic model there for illustrative purposes here. 
The shape of the angular power spectrum is essentially a reflection of the projected linear power spectrum over the redshift distribution of background sources. Since first stars are more likely to be found in the similar redshift range as of the neutral Hydrogen, which contributes to the 21 cm background,
we expect the shape of the anisotropy clustering power spectrum of the 21 cm background to remain the same.
Due to differences in emissivity, of course, the clustering power spectrum will have a different amplitude in this case.

In Fig.~3, we also show the angular power spectrum of spatial fluctuations in the 
background diffuse emission after gravitational lensing. While the lensing effect results in a transfer of
power from arcminute angular scales, where the angular power spectrum peaks, to tens of degree angular scales, where the power spectrum is
rapidly rising, there is no substantial modification to the angular power spectrum at arcminute angular scales. 
In comparison, as shown in Fig.~2, the lensing effect on
CMB anisotropies results in a transfer of power from degree angular scales, where the power spectrum peaks in this case, to arcminute angular scales
such that there is a substantial enhancement of power along the damping tail by more than two orders  of magnitude. 
The lack of lensing modification to the clustering power spectrum of diffuse backgrounds can be understood based on Eq.~15
and results from that fact that clustering at arcminute scales is smooth and results from the collective emission from a large number of sources such that
under gravitational lensing, though there may be modifications to the distribution of these sources at arcsecond angular scales, the distribution
at arcminute angular scales remain the same. If any differences exist, these are at the
level substantially below a percent. The lack of a substantial difference in the angular power spectrum before and after gravitational lensing
suggests that statistics that primarily target two-point clustering information will not be useful for lensing studies. For example, the variance
related to the distribution of these background fluctuations will not be substantially varying and we suggest that the variance measurements may not
be ideal. The lack of a major difference also limits the extent to which lensing information can be extracted from non-Gaussian statistics, such as those
related to the trispectrum that have been considered with respect lensing studies with CMB. 

As discussed earlier, our calculation on the lensing effect on background clustering power spectrum may be incomplete. Since the typical deflection angle
is of order an arcminute and there is substantial clustering power at arcminute scales, it is necessary that one properly accounts for 
higher order corrections to the previous calculations. We have attempted this numerically and show the first and second order contributions 
in Fig.~3. Unlike the case with CMB anisotropies, where the second order contribution was lower than the first order term, at angular scales
of arcminutes and below, we find that this to be no longer the case here. In the multipole range of interest, however, we expect there to be no
substantial contribution from the second and other higher order terms, though if we were to study the lensing corrections to the clustering power spectrum
at arcsecond angular scales, then it may be necessary to perform an improved calculation. We restrict to arcminute scale clustering here
since such clustering may easily be detectable in near future. On 
the other hand, even if clustering at arcsecond angular scales is detected, we expect at such angular scales,
 the shot-noise associated with finite density of sources that contribute to the background emission to be the dominant contribution to clustering.
In Fig.~3, for illustration, we show the expected shot-noise related to the IR background generated by first stars and galaxies following
calculations in Cooray et al (2003).

While we have specifically discussed the lensing modification to the IR background generated by first stars, our conclusions apply to clustering aspects 
of all other backgrounds. As argued earlier, for backgrounds at the same redshift range, one expects a similar shape to the angular power spectrum
though the amplitude is different for obvious reasons. Our calculation related to the lensing modification of the background fluctuation power spectrum is
independent of its amplitude and the difference between the lensed and unlensed power spectra scales accordingly such that the fractional difference between
the two remain the same. The only differences are, however, associated with secondary considerations such as the extent to which shot-noise
related to the finite density of sources that contribute to each of these backgrounds may become a source of confusion noise for lensing reconstruction.

\begin{figure}[t]
\centerline{
\psfig{file=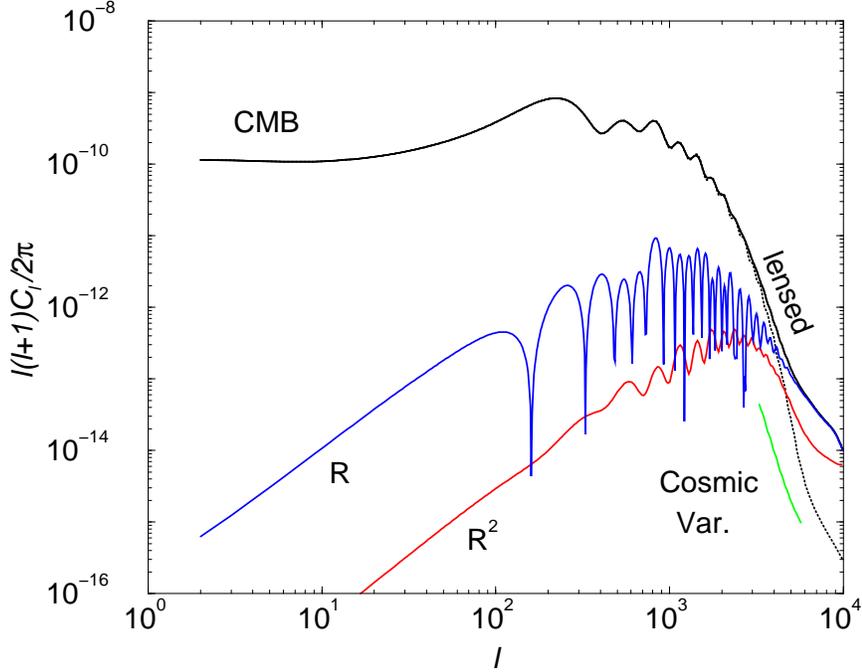,width=4.5in,angle=-90}}
\caption{The lensed CMB temperature anisotropy power spectrum. The curves labeled 'R' and 'R$^2$' show the lensing calculation up to the
first- and second-order in $C_l^{\phi}$, respectively. Note that typical calculations in the literature involve the first order calculation,
while the next-order correction considered here, the R$^2$ contribution, is at the level of 30\% and below and can be ignored for most practical purposes;
for precision cosmology, where lensing information is used to extract cosmological parameters, such higher order corrections must be accounted for and
we illustrate this why by showing the cosmic variance limit (line labled 'Cosmic Var.') of the power spectrum measurements at  each of these multipoles
with a full-sky experiment.}
\label{fig:cmb}
\end{figure}

\begin{figure}[t]
\centerline{
\psfig{file=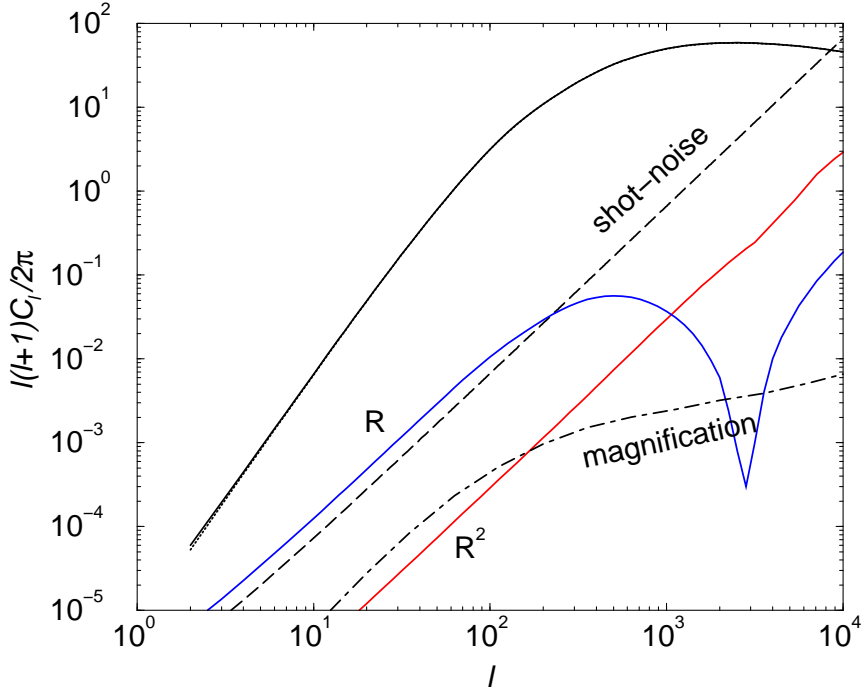,width=4.5in,angle=-90}}
\caption{The clustering power spectrum of a diffuse background at near-IR wavelengths due to first stars and galaxies and at redshifts between 10 and 30.  The dotted line shows the intrinsic angular power spectrum, while the correction related to gravitational lensing is shown with a 
dashed line. The solid line is the angular power spectrum one observes today; at large angular scales, this power spectrum is few percent
larger than the intrinsic power spectrum while the differences are insignificant at arcminute angular scales.
The dashed line shows an estimate on the shot-noise contribution to the clustering signal when integrated over the finite number density
of background sources; for diffuse emissions such as 21cm line from the general IGM prior to reionization, we expect this shot-noise to be
significantly lower or non-existent. The curves labeled 'R' and 'R$^2$' show the first and second-order contributions related to the lensing
calculation. The curve labeled 'magnification' is an estimate on the magnification-related 
correction to the power spectrum when a finite number density of sources is 
involved and the background is studied after removing certain resolved bright sources.}
\label{fig:clirb}
\end{figure}

\begin{figure}[t]
\centerline{
\psfig{file=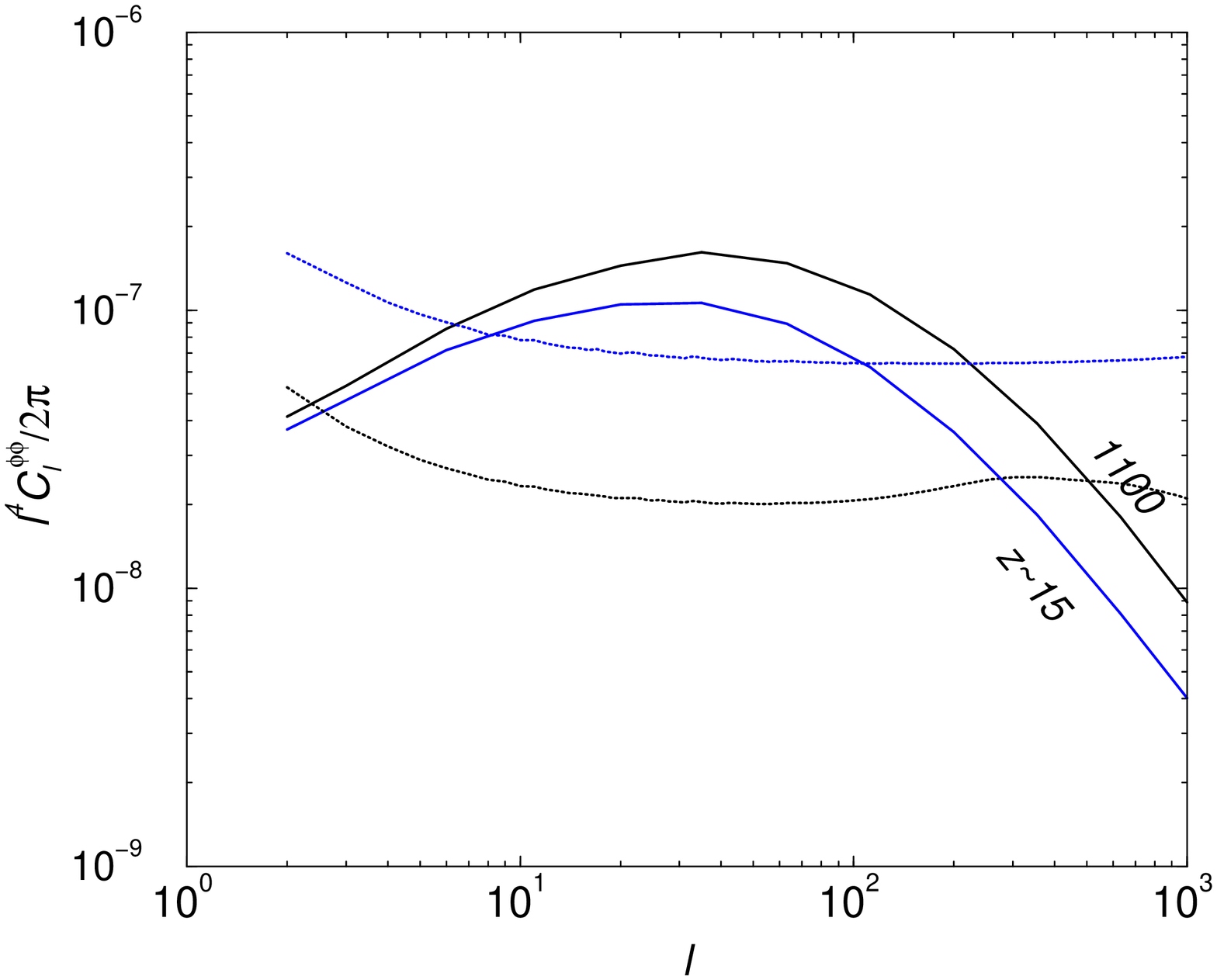,width=4.5in,angle=-90}}
\caption{The extraction of lensing information from clustering data related to the diffuse background. The solid lines show the power spectrum of
projected potentials while the dotted line show the expected noise associated  with the lensing reconstruction. Here, we show the case at $z \sim 15$,
based on, for example, IR background from first stars and galaxies, or 21 cm emission from neutral Hydrogen, while, for comparison, we also show the
case with CMB using, for example,e arcminute scale resolution data expected from a mission such as CMBpol. In comparison, CMB temperature fluctuations allow 
a significantly better reconstruction related to gravitational lensing than low redshift backgrounds and is primarily due to the difference
in the fluctuation pattern on the sky or the clustering power spectrum. The presence of features and the lack of power at arcminute scales in CMB
anisotropies enhance the lensing modification, while in the case of diffuse backgrounds, the lensing modifications are minor and the generated 
non-Gaussianities are no significant.}
\label{fig:error}
\end{figure}

\subsection{Lensing Reconstruction from the Diffuse Emission}

As a first approach, one can consider a
lensing reconstruction using diffuse background anisotropy data similar to that suggested for CMB.
The idea behind here is the presence of a significant non-Gaussian signal in the form of a trispectrum, and an approach to
probe this non-Gaussianity based on quadratic statistics such the power spectrum of CMB-gradients (Zaldarriaga \& Seljak 1999; Hu 2001; Hirata \& Seljak 2003). Here, we make use of the discussion in Cooray \& Kesden (2003; see, also, Hu 2001; Hu \& Okamoto 2002). 
Under gravitational lensing, the trispectrum for diffuse backgrounds takes the form of
\begin{eqnarray}
&&\tilde T^I(\bfl_1,\bfl_2,\bfl_3,\bfl_4) = -\int d\rad C_{l_3}^{\cmb} C_{l_4}^{\cmb} \Big[
C^\pp_{|\vecl_1+\vecl_3|}(\rad) (\vecla +\veclc) \cdot \veclc (\vecla + \veclc) \cdot \vecld \nonumber \\
&& \quad + C^\pp_{|\vecl_2+\vecl_3|}(\rad) (\veclb +\veclc) \cdot \veclc (\veclb +\veclc)  \cdot \vecld \Big]
+ \, {\rm Perm.} \, ,
\label{eqn:trilens}
\end{eqnarray}
where the permutations contain 5 additional terms with the replacement of
$(l_3,l_4)$ pair by other combination of pairs. The derivation related to Eq.~\ref{eqn:trilens} can be found in Cooray \& Kesden (2003), though, 
we have modified it slightly to account for the finite width in the background source distribution by including an explicit integral over the radial
distribution. The lensing extraction make use of this trispectrum based on the fact that it is simply proportional to $C_l^{\phi\phi}$.

In practice, one makes an estimator for the deflection angle 
based on quadratic statistics. In the present case, when compared to CMB, note that the one estimates a weighted average of the deflection angle
over the redshift distribution of sources that contribute to the diffuse emission. The estimator is such that, in Fourier space,
one is convolving two filtered versions of the background intensity fluctuations
\begin{equation}
\hat{I}^2(\vecl) = \int \frac{d^2\vecl_1}{(2\pi)^2} W(\vecl,\vecl_1) \tilde I(\vecl) \tilde I(\vecl-\vecl_1) \, ,
\end{equation}
where $W$ is a filter that is designed to optimally extract information related to the weak lensing effect such that the
power spectrum of $\hat{I}^2(\vecl)$ returns $\langle C_l^\phi \rangle$ where
\begin{equation}
\langle C_l^\phi \rangle = \frac{\int dr C_l^\phi(r) C_l^\cmb}{\int dr C_l^\cmb} = \int dr n^2_s(\rad) C_l^\phi(\rad) \, ,
\end{equation}
where $n_s(\rad)$ is the normalized distribution of background sources that contribute to the diffuse emission. 
The filter, $W(\vecl,\vecl_1)$, can be represented by typical quadratic combinations considered in the literature, such as gradients (e.g., Zaldarriaga \& Seljak 1999).
To extract all information, following the case in CMB anisotropies, the optimized filter to extract lensing information takes the form of
\begin{equation}
W(\vecl,\vecl_1) = \frac{[\vecl \cdot \vecl_1 C_{l_1} + \vecl \cdot (\vecl - \vecl_1) C_{|\vecl-\vecl_1|}]}{2C_{l_1}^{\rm tot}C_{|\vecl-\vecl_1|}^{\rm tot}} \, ,
\end{equation}
such that when one constructs $N_l^2 \langle \hat{I}^2(\vecl) \hat{I}^2(\vecl') \rangle$ one measures $\langle C_l^\phi \rangle$ with a
noise contribution given by
\begin{equation}
N_l = \left[\int \frac{d^2\vecl_1}{(2\pi)^2} \frac{[\vecl \cdot \vecl_1 C_{l_1} + \vecl \cdot (\vecl - \vecl_1) C_{|\vecl-\vecl_1|}]^2}{2C_{l_1}^{\rm tot}C_{|\vecl-\vecl_1|}^{\rm tot}} \right]^{-1} \, .
\end{equation}
Here $C_l^{\rm tot}$ is the total contribution to the angular power spectrum of the diffuse background map and includes
contributions such as $C_l^{\rm tot} = \tilde C_l + C_l^{\rm shot} + C_l^{\rm mag} + C_l^{\rm noise}$, which are related to the
unlensed clustering power spectrum including any shot-noise, $\tilde C_l + C_l^{\rm shot}$, any corrections related to the magnification, if again
there is a finite number density of sources involved $C_l^{\rm mag}$ and instrumental noise, $C_l^{\rm noise}$.

The shot-noise associated with a finite density of sources that determine the background emission can be
estimated through number counts, $dN/dS$, of the contributing
sources, as a function of flux $S$, and can be written as
\begin{equation}
C_l^{\rm shot} = \int_{0}^{S_{\rm cut}} S^2 \frac{dN}{dS} \; dS \, ,
\end{equation}
where $S_{\rm cut}$ is the flux cut off value  related to the removal of resolved sources. This shot-noise
acts as a source of noise for lensing reconstruction.  Unlike the calculation related to the clustering spectrum, where the average
emissivity is the only unknown, the shot-noise depends strongly on detailed aspects of source number counts. Since one weighs
by a factor $S^2$, the shot-noise is more sensitive to rare objects which are brighter. Since there are many uncertainties related to establishing
the source number counts precisely, we loosely estimate this following Cooray et al. (2003) and show this  estimate in Fig.~3 for the case
of IR-background generated by first stars and galaxies. Note that for ``truly'' diffuse emissions, such as the 21 cm rest-wavelength background 
from general IGM and CMB, this is not a concern. 

Note that, so far, we have considered the case where the clustering analysis takes
in to account all contributions to the background. In some situations, such as in cases where some part of the background is due to resolved
sources in data, one can study clustering properties by removing such sources from the analysis. While we do not
expect this to be the case either with near-IR background from first objects or in 21 cm emission, most dusty galaxies that contribute to 
the far-IR background  can be resolved and removed. In such a  case, the clustering power spectrum of the residual emission 
contain a correction associated with lensing magnification (Tegmark \& Villumsen 1997) and can be written as
\begin{equation}
C_l^{\rm shot} = l^4 \langle C_l^{\phi\phi} \rangle S_{\rm cut}^4 \left[\frac{dN}{dS}\right]^2_{S_{\rm cut}} \, .
\end{equation}
This noise contribution 
depends on the number counts at the flux-cut off. If this contribution can be identified, it alone may be used for a lensing study
though the modification to the power spectrum is expected to be small.
Though we make an estimate of the contribution in Fig.~3 (curve labeled 'magnification'),
this estimate should also be considered as highly uncertain since number counts of sources that contribute to the background are not
well defined in the context of backgrounds we have primarily considered. This effect is again only present in the case of a finite density of sources
and is not a concern when, for example, all sources that contribute to the background have the same flux.

In addition to effects related to a finite density of sources that lead to a diffuse background, the presence of instrumental noise
also affect the lensing reconstruction. This noise is given by
\begin{eqnarray} \label{E:noise}
C_l^{\n} = f_{\rm sky} w^{-1} e^{l^2 \sigma_{b}^2} \, ,
\end{eqnarray}
when $f_{\rm sky}$ is the fraction of sky surveyed, $w^{-1} = 4\pi \sigma^2_{\rm pix}/N_{\rm pix}$
is the variance per unit are on the sky with an individual pixel noise variance of $\sigma^2_{\rm pix}$
with $N_{\rm pix}$ pixels, and $\sigma_b$ is the effective beam-width of the instrument. 

In Fig.~4, we illustrate the extent to which CMB and low redshift diffuse backgrounds can be used for a lensing reconstruction by concentrating on
the expected level of noise associated with the reconstructed potential power spectrum. For simplicity, here we consider arcminute scale resolution
experiments, but set the noise level to be ten orders of magnitude below the clustering power spectrum such that at multipoles
of few hundred, the noise-level is insignificant. This is done to consider the full extent to which diffuse backgrounds can be used for lensing
studies based on clustering information and to compare with CMB related lensing studies. As shown in Fig.~4, CMB data allow
an order of magnitude or better reconstruction of the projected lensing potential power spectrum when compared to the case with diffuse backgrounds.
This is due reasons we have already mentioned and involve the fact that the non-Gaussianity generated in CMB data, by the nature of its fluctuation
pattern, is more significant than the non-Gaussianity generated by gravitational lensing in the low redshift diffuse emission. While fluctuations in the diffuse
emission may allow a reconstruction of the lensing power spectrum with a cumulative signal-to-noise ratio of order few hundred, 
this should be compared to at least an order of magnitude better reconstruction with CMB with the same angular resolution and fractionally
the same noise level. Unlike low redshift backgrounds, CMB also has the advantage that a lensing reconstruction can be considered from
polarization information in addition to temperature anisotropies we have only considered here (e.g., Guzik et al. 2001; Hu \& Okamoto 2002). 

In addition to problems related to the reconstruction, there are also reasons that can complicate a simple interpretation of the
reconstructed lensing power spectrum. For example, in addition to the lensing contributions we have discussed,
the lensed clustering power spectrum also contains additional 
contributions related to the fact that the source distribution that contribute to low redshift background is broad and these lensing
potentials can be correlated with the diffuse emission itself. We write the dominant contribution associated with such a cross-correlated component as
\begin{eqnarray}
C_l^{\rm conf}=  \intl{1} \int d\rad_1  C_{l_1}^{\phi-\delta I}(\rad_1)  \int d\rad_2 C_{| \vecl - \vecl_1|}^{\phi-\delta I}(\rad_2) [(\vecl - \vecl_1)\cdot \vecl_1]^2 \, .
\end{eqnarray}
This term does not exist in the case of CMB since the potentials that deflect CMB photons are
disjoint in redshift space from last scattering surface where most 
CMB fluctuations are generated. In the case of diffuse backgrounds, if the source distribution is broad, this term can be interpreted simply as the
contribution due to lensing potentials at the low end of the source distribution that is responsible for the lensing
of sources at the other end of the distribution. Assuming that both potentials and the sources of background light anisotropy trace the 
same power spectrum $P(k)$, with a bias factor $b(k,\rad_2)$ that accounts for any departures from this assumption,
this cross power spectrum can be written as
\begin{eqnarray}
C_l^{\phi-\delta I}(\rad_s) &=& \frac{2}{\pi} \int k^2\, dk P(k)
\int_0^{\rad_s} d\rad W^\lens(k,\rad_1) j_l(k\rad_1) \int_0^{\infty} d\rad_2 b(k,\rad_2) b(k,\rad_2) j_l(k\rad_2) \, .
\label{eqn:cross}
\end{eqnarray}
A proper accounting of effects such as this will be a necessary aspect of any study that attempt to use background source clustering for
lensing purposes.

While a reconstruction with CMB is better, note that one reconstructs the potential power spectrum out to a redshift of 1100. Thus, low redshift 
backgrounds
are still useful for cosmological studies since they allow a reconstruction out to redshifts of 15 to 30 or so and will bridge the gap between
low redshift lensing studies, based on galaxy shapes, and eventual lensing studies with higher resolution CMB data  in the near future.
In this respect, we suggest that one should concentrate further on the aspects of lensing related to diffuse backgrounds and how they can be exploited.
As we have discussed, the use of non-Gaussian information for a lensing reconstruction with low redshift backgrounds, 
in the same manner lensing studies with CMB are proposed, is not helpful, though an attempt is still encouraged. To extract lensing information,
one should concentrate on the distribution of sources, that contribute to the background emission, at arcsecond angular scales.
One approach would be to use the local anisotropy of the correlation function to obtain an estimate for the shear (van Waerbeke et al. 1997); 
this comes from the fact while
clustering properties of the background is expected to isotropic, in addition to non-Gaussianities, gravitational lensing also modify the
clustering such that, locally, an anisotropic correlation function is generated. The quadrupole moment of the local correlation function captures this
anisotropy and provides an estimate of shear just as ellipticity is used to estimate lensing shear. Another approach is to use
variations in number counts due to lensing magnification related effects (Jain 2002), though this requires backgrounds where sources counts are
easily modified by magnification due to the nature of source count distribution as a function of redshift. While we have primarily discussed the use
of arcminute scale clustering information and the potential presence of non-Gaussianities for a lensing reconstruction, in a future paper,
we will return to the issue of extracting lensing shear from small angular scale clustering properties.

\section{Summary}

The current weak lensing measurements of the large scale structure are mostly related to statistical
study of background galaxy ellipticities. We consider a possibility to extend lensing studies with
intrinsically unresolved sources and suggest that clustering properties, at arcminute angular scales, of either
the point sources or the integrated diffuse emission from these point sources can be used
for a lensing reconstruction. This is analogous to  techniques now developed to extract lensing information from
cosmic microwave background  (CMB) temperature and polarization fluctuations and relies on the presence of a significant non-Gaussian signal
in the background generated by the gravitational lensing modifications to the fluctuation pattern.
Contrary to CMB, however, the lensing modification to the diffuse background clustering is minimal and results from the 
lack of distinct features in the fluctuation pattern. In the case of CMB, a  distinct signature or pattern  exist through the
acoustic nature and the lack of intrinsic power at arcminute angular scales where the damping-tail is found.
We have discussed the extent to which projected matter power spectrum be reconstructed with images of the diffuse background generated by
first stars and galaxies in near-IR wavelengths and the background related to 21 cm emission by neutral gas in the general
intergalactic medium prior to reionization and have shown this to be significantly below the level of signal-to-noise ratio one can
reach with, in relative terms, CMB data. This is due to the lack of significant non-Gaussianities in the diffuse emission that is
generated by gravitational lensing modifications. Though limited, lensing studies with diffuse backgrounds will only be the
way to extend the reconstruction of the integrated matter power spectrum out to redshifts
of 15 to 30 and to bridge the gap between current and upcoming lensing studies with sources at redshifts between 1 and 2
and planned weak lensing studies with CMB out to the last scattering surface at a redshift of 1100.
As ways to improve, we suggest further studying a potential lensing reconstruction methods based on shear estimation from the local
anisotropy of the correlation function of the sources that contribute to diffuse  backgrounds and statistics that are optimized to extract
information related to modifications associated with lensing magnification.

\section*{Acknowledgments}
The author thanks Roger Blandford, Marc Kamionkowski, and Michael Kesden for useful discussions and collaborative work.
This work was supported in part by DoE DE-FG03-92-ER40701 and a senior research fellowship from the Sherman Fairchild Foundation.



\end{document}